\documentclass[%
 reprint,
superscriptaddress,
 amsmath,amssymb,
 aps,
prb,
]{revtex4-2}

\usepackage{graphicx}
\usepackage{dcolumn}
\usepackage{bm}
\usepackage[hidelinks]{hyperref}
\usepackage{subcaption}
\usepackage{xcolor}
\usepackage{multirow}

\newcommand{\orcid}[1]{\href{https://orcid.org/#1}{\includegraphics[scale=0.5]{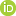}}}

\begin{document}

\preprint{APS/123-QED}

\title{On the experimental properties of the TS defect in 4H-SiC}

\author{Johannes A.\ F.\ Lehmeyer\orcid{0000-0003-2041-9987}}%

 \affiliation{Lehrstuhl für Angewandte Physik, Friedrich-Alexander-Universität Erlangen-Nürnberg, Staudststr.\ 7, 91058 Erlangen, Germany.
}

\author{Alexander D.\ Fuchs\orcid{0000-0003-1896-9242}}%
 \affiliation{Lehrstuhl für Angewandte Physik, Friedrich-Alexander-Universität Erlangen-Nürnberg, Staudststr.\ 7, 91058 Erlangen, Germany.
}

\author{Zhengming Li}%
 \affiliation{Lehrstuhl für Angewandte Physik, Friedrich-Alexander-Universität Erlangen-Nürnberg, Staudststr.\ 7, 91058 Erlangen, Germany.
}

\author{Titus Bornträger}%
 \affiliation{Lehrstuhl für Angewandte Physik, Friedrich-Alexander-Universität Erlangen-Nürnberg, Staudststr.\ 7, 91058 Erlangen, Germany.
}

\author{Fabio Candolfi\orcid{0009-0004-1187-590X}}%
 \affiliation{Lehrstuhl für Angewandte Physik, Friedrich-Alexander-Universität Erlangen-Nürnberg, Staudststr.\ 7, 91058 Erlangen, Germany.
}

\author{Maximilian Schober\orcid{0000-0001-9788-3302}}
 \affiliation{Institute for Theoretical Physics, Johannes Kepler University Linz, Altenbergerstr. 69, A-4040 Linz,
Austria.}


\author{Marcus Fischer\orcid{0000-0001-6679-0509}}
 \affiliation{Erlangen Center for Interface Research and Catalysis (ECRC), Friedrich-Alexander-Universität Erlangen-Nürnberg, Egerlandtsr. 3, 91058 Erlangen, Germany.}

\author{Martin Hartmann\orcid{0000-0003-1156-6264}}
 \affiliation{Erlangen Center for Interface Research and Catalysis (ECRC), Friedrich-Alexander-Universität Erlangen-Nürnberg, Egerlandtsr. 3, 91058 Erlangen, Germany.}

\author{Elke Neu\orcid{0000-0003-1904-3206}}
 \affiliation{Rheinland-Pfälzische Technische Universität Kaiserslautern-Landau, Department of Physics, Erwin Schrödinger Strasse, D-67663 Kaiserslautern, Germany.}

\author{Michel Bockstedte\orcid{0000-0001-5720-4010}}
 \affiliation{Institute for Theoretical Physics, Johannes Kepler University Linz, Altenbergerstr. 69, A-4040 Linz,
Austria.}

\author{Michael Krieger\orcid{0000-0003-1480-9161}}%
 \affiliation{Lehrstuhl für Angewandte Physik, Friedrich-Alexander-Universität Erlangen-Nürnberg, Staudststr.\ 7, 91058 Erlangen, Germany.
}

\author{Heiko B.\ Weber\orcid{0000-0002-6403-9022}}%
\email{heiko.weber@fau.de}
 \affiliation{Lehrstuhl für Angewandte Physik, Friedrich-Alexander-Universität Erlangen-Nürnberg, Staudststr.\ 7, 91058 Erlangen, Germany.
}

\date{\today}

\begin{abstract}
When annealing a 4H silicon carbide (SiC) crystal, a sequence of optically active defect centers occurs among which the TS center is a prominent example.
Here, we present low-temperature photoluminescence analyses on the single defect level. They reveal that the three occurring spectral signatures TS1, TS2 and TS3 originate from one single defect.
Their polarization dependences expose three different crystallographic orientations in the basal plane, which relate to the projections of the nearest neighbor directions.
Accordingly, we find a three-fold level-splitting in ensemble studies, when applying mechanical strain. This dependency is quantitatively calibrated.
A complementary electrical measurement, deep level transient spectroscopy, reveals a charge transition level of the TS defect at 0.6\,eV above the valence band.
For a future identification, this accurate characterization of its optical and electronic properties along with their response to mechanical strain is a milestone.
\end{abstract}

\maketitle
\section{\label{sec:intro}Introduction}

Wide bandgap semiconductors provide the opportunity to study atomic physics and spin physics in a technically controllable solid state environment. Conceptually, a single defect in an otherwise perfect crystal closely resembles an atom or molecule in empty space, provided the relevant states are deep within the bandgap.
The most frequently studied point defect in this respect is the NV-center in diamond, which is used as single photon source, quantum sensor, quantum communication node, etc. \cite{doherty:11, lenzini:18}.
It would be desirable to embed such a spin-carrying point defect in an electrically controlled environment, such that both photonic quantum technology with its optical methodology and the powerful classical electronic methodology can monolithically be combined \cite[Chapter 8]{Castelletto2015}. Diamond as a host material is not ideally suited for such monolithic integration, because, in particular, n-doping capabilities are insufficient \cite{wort:08}.

Within the mindset of integrating quantum technologies and semiconductor electronics, silicon carbide (SiC) is ideally suited \cite{Atature2018, Lukin2020}.
It has inherited from silicon the capabilities as a tractable semiconductor including p- and n-type doping. Today, a powerful and mature device technology is available for the hexagonal polytype 4H-SiC \cite{She2017}.
Moreover, epitaxial graphene on 4H-SiC is an optically transparent yet metallic material that can be monolithically integrated with SiC \cite{hertel:12, Ruhl2020}.
As to quantum technology, research in SiC focuses on the silicon vacancy V$_\text{Si}$ and the divacancy V$_\text{Si}$V$_\text{C}$, with promising capabilities as qubits, allowing for quantum sensing, quantum computing and quantum communication applications \cite{janzen:09, kraus:14, koehl:11, son:06}.

Here, we present a research on a less explored point defect in 4H SiC, the TS defect. We will investigate its optical and electronic properties in response to optical, mechanical and electrical stimuli. The TS defect is created by annealing irradiated crystals at very high temperatures, such that it can reside in an otherwise perfect crystalline environment, which would be ideal for quantum applications \cite{ruehl:18}. Further, as early ensemble measurements have demonstrated, the TS has a very strong spectral shift upon application of mechanical strain or electrical fields, which makes it attractive as point-like sensor \cite{ruehl:21}.

Based on these ensemble measurements, an earlier tentative model, namely the carbon di-vacancy-antisite complex (V$_\mathrm{C}$C$_\mathrm{Si}$V$_\mathrm{C}$) \cite{schober:23}, was partially supported by an erroneous assignment of dipole directions that we corrected only recently \cite{ruehl:21}.
The present study involves single TS centers and provides further insights that will help to find a microscopic model.


The TS lines become apparent in low temperature photoluminescence (PL) when SiC samples with intrinsic damage (e.g.\ created by proton implantation) are annealed. Immediately after irradiation the silicon vacancy V$_\text{Si}$ is predominant. Upon annealing, V$_\text{Si}$ disappears and instead the divacancy V$_\text{Si}$V$_\text{C}$ and the antisite-vacancy C$_\text{Si}$V$_\text{C}$ occur. At even higher temperatures (around 1200°C), both vanish and the TS color center becomes apparent in the otherwise clean crystal, i.e.\ no further PL signatures are remaining in the investigated spectral window \cite{ruehl:18, Kobayashi2022}. Its signature is a very sharp line at 769\,nm (TS1), accompanied by two satellite lines at 812\,nm and 813\,nm (TS2, TS3). The zero-phonon line of TS1 is unusually pronounced whereas the phonon side-band is weak, resulting in a Debye-Waller factor of about 45\% \cite{ruehl:18}.\\
Initial measurements of the PL of TS ensembles indicated that they are very susceptible to perturbations. In particular, the ensemble splits in three different PL-lines in strained areas, for example next to a scratch \cite{ruehl:21}. Using epitaxial graphene electrodes, the TS shows also a strong response in Stark-effect studies, where static electric fields were applied in various crystal orientations. It was identified that the dipole vector projection in the basal plane is aligned with that of the vectors connecting nearest neighbors (Si-C) in SiC's crystal structure \cite{ruehl:21}.


\section{\label{sec:block1}Experimental procedure}
\subsection{Sample Fabrication}
For measurements of single defects, ensembles under strain and spin investigations, Wolfspeed high-purity semi-insulating (HPSI) 4H-SiC on-axis ($0.08^\circ$ vs c-axis) wafers were used. These samples were cleaned with acetone in an ultrasonic bath. Then organic contaminations were removed with a Piranha acid (three parts of 98\,\% sulfuric acid and 1 part of 30\,\% hydrogen peroxide).

As the TS defect is present in smallest concentrations in most 4H-SiC samples even without implantation, to reduce their density, the HPSI samples for single defect measurements were annealed to 1700°C in 900\,mbar Ar-atmosphere for 30.
To obtain a smooth surface, the C-face of the sample used for the single defect measurements was mechanically polished after annealing.

The samples for the strain dependent measurements were cut into rods of $8\times 0.5$\,mm$^2$. The rods were oriented along three crystallographic directions: $[1\overline{1}00]$, $[11\overline{2}0]$, and 45° in between the other two. To create the TS color center ensemble, the samples were proton irradiated with an energy of 350\,keV and a dose of $10^{15}$\,cm$^{-2}$ and subsequently annealed at 1200°C under 900\,mbar Ar-atmosphere for 30 minutes to maximize the number TS defects and reduce the amount of other defects.

For the spin investigation, samples were proton irradiated with a dose of $10^{15}$\,cm$^{-2}$ at an energy of 350\,keV and subsequently annealed at 1200°C under 900\,mbar Ar-atmosphere for 30 minutes. 

The DLTS samples were cut from commercially available p-type wafers with a nominal Al doping of $7\cdot10^{15}$\,cm$^{-3}$ in the 8\,$\mu$m thick top epitaxial layer. After an RCA-clean, 150\,nm of Al were sputtered onto the backside of the sample, and subsequently annealed at 950°C for 3 minutes to form ohmic contacts. The samples were proton irradiated with a dose of $1\cdot10^{12}$\,cm$^{-2}$ and an energy of 350\,keV. The annealing at temperatures $T_\mathrm{A} < 1000$°C was performed in vacuum, for $T_\mathrm{A} > 1000$°C it was performed under 900\,mbar Ar-atmosphere.
Implantation and annealing were done after creating the backside contact for the samples with $T_\mathrm{A} < 1000$°C and vice versa for the ones with $T_\mathrm{A} > 1000$°C.
As a last step, 150\,nm Ni Schottky contacts were deposited on the front side.

\subsection{Experimental Setups}
The optical measurements were performed in custom-built confocal microscope setups. For the strain measurements, an objective with a numerical aperture of 0.45, for the single defect measurements an objective with a numerical aperture of 0.65 and in both cases a 50\,$\mu$m pinhole were used. A 532\,nm laser was used for optical excitation. The spectra were recorded with an Andor Shamrock 500i spectrometer and an Andor Newton 920 CCD.
The single defect measurements were performed at 4\,K, the strain measurements at 100\,K.

The optical measurements under mechanical strain were performed in a squeezable nanojunction setup (without a second sample) \cite{Popp2021}, i.e.\ the sample is pressed from below with a mandrel and held tightly at the sides.
Similar measurements have already been presented in conference proceedings, where also a view of the setup is provided \cite{lehmeyer:23}.

For the DLTS measurements, the system PhysTech FT1020 was used. The measurement temperature was controlled with a PID controller using liquid air as a coolant and programmable power supply for heating. Period widths of 16, 32 and 64\,ms were chosen; the pulse width was 0.1\,ms.


\section{Results}

\subsection{Polarization of single TS defects}
\begin{figure*}
    \centering
    \begin{subfigure}[b]{0.49\textwidth}
        \centering
        \includegraphics[width=\textwidth]{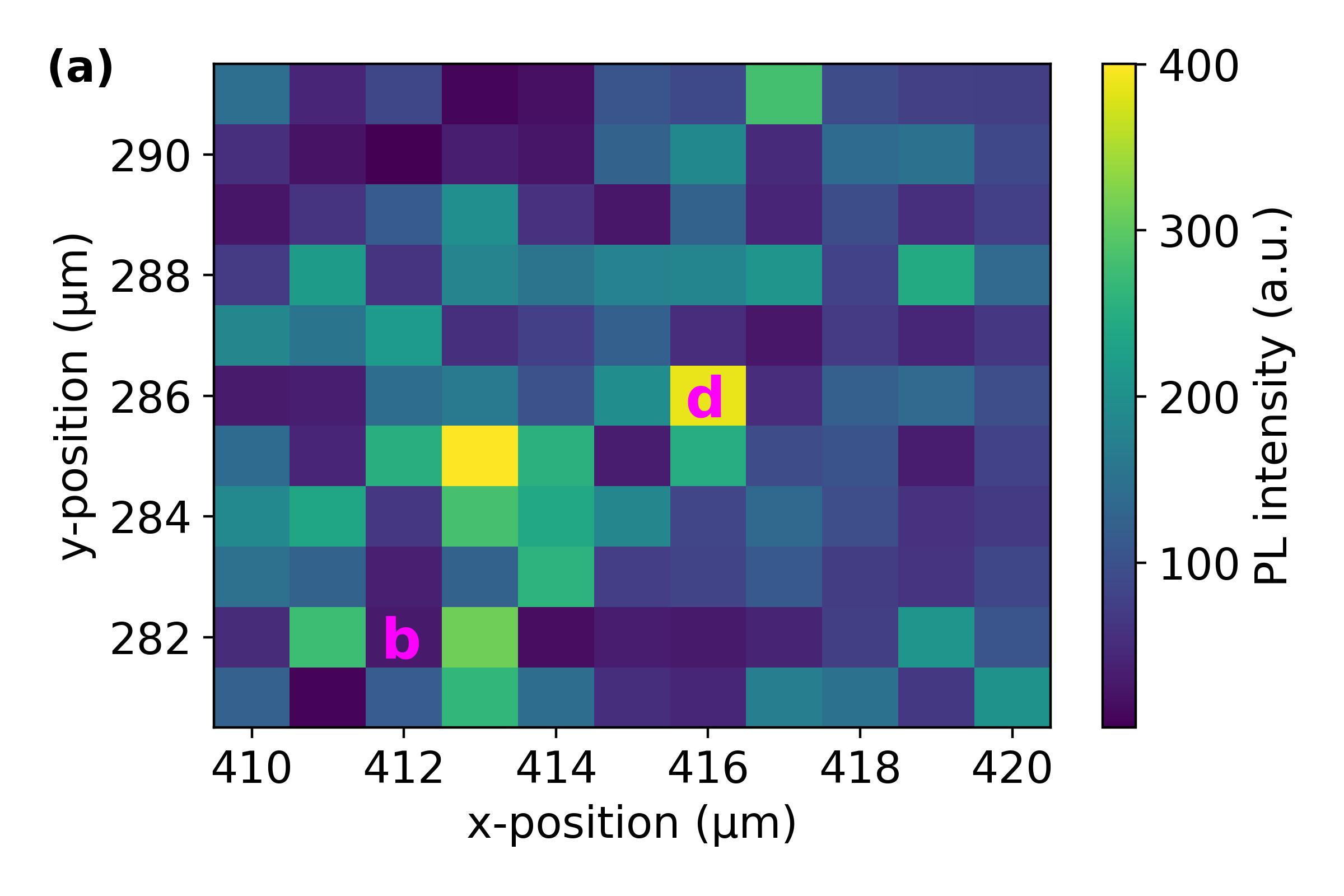}
    \end{subfigure}
    \hfill
    \begin{subfigure}[b]{0.49\textwidth}
        \centering
        \includegraphics[width=\textwidth]{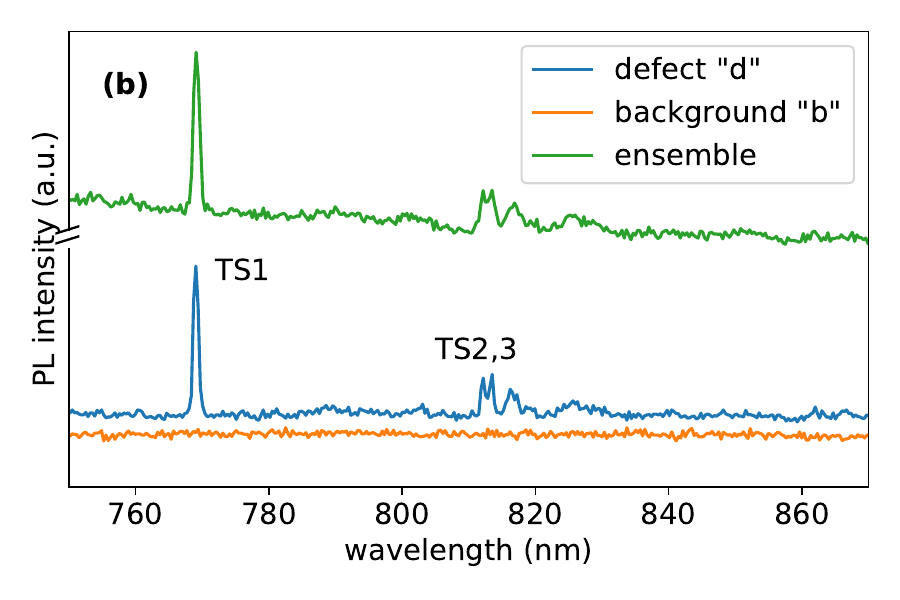}
    \end{subfigure}\\
    \begin{subfigure}[b]{0.32\textwidth}
        \centering
        \includegraphics[width=\textwidth]{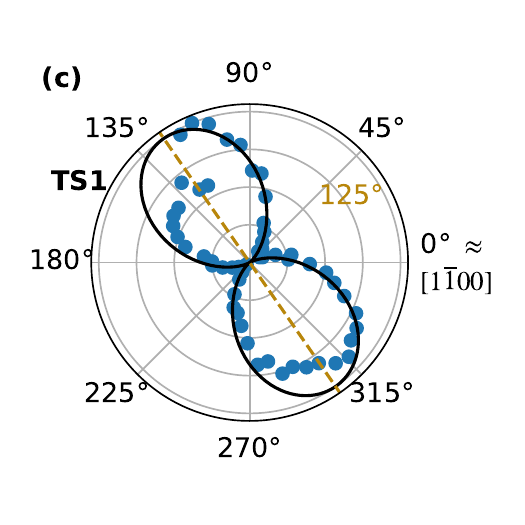}
    \end{subfigure}
    \hfill
    \begin{subfigure}[b]{0.32\textwidth}
        \centering
        \includegraphics[width=\textwidth]{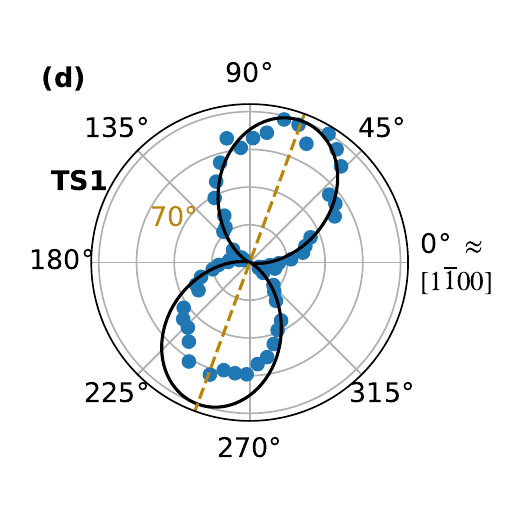}
    \end{subfigure}
    \hfill
    \begin{subfigure}[b]{0.32\textwidth}
        \centering
        \includegraphics[width=\textwidth]{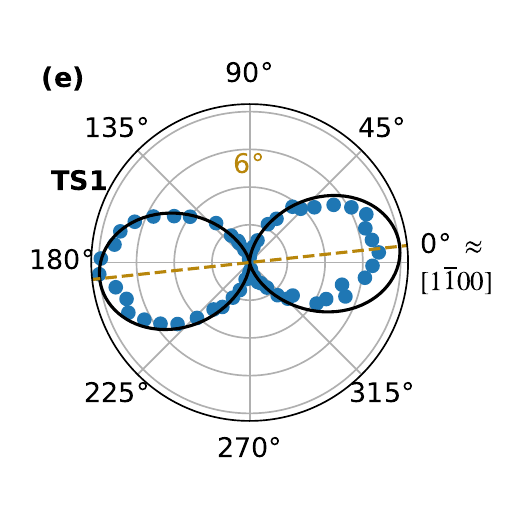}
    \end{subfigure}\\
    \vspace{-0.7cm}
    \begin{subfigure}[b]{0.32\textwidth}
        \centering
        \includegraphics[width=\textwidth]{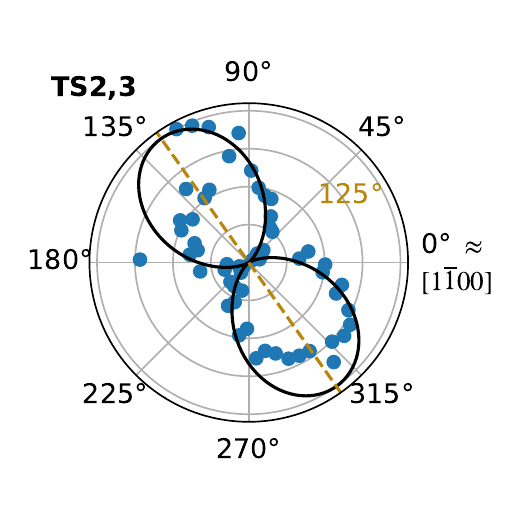}
    \end{subfigure}
    \hfill
    \begin{subfigure}[b]{0.32\textwidth}
        \centering

        \includegraphics[width=\textwidth]{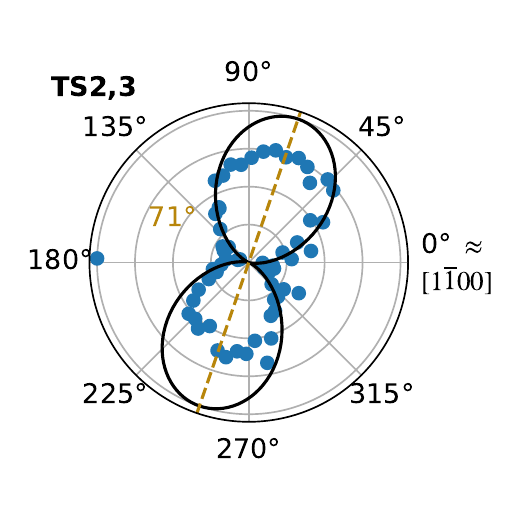}
    \end{subfigure}
    \hfill
    \begin{subfigure}[b]{0.32\textwidth}
        \centering
        \includegraphics[width=\textwidth]{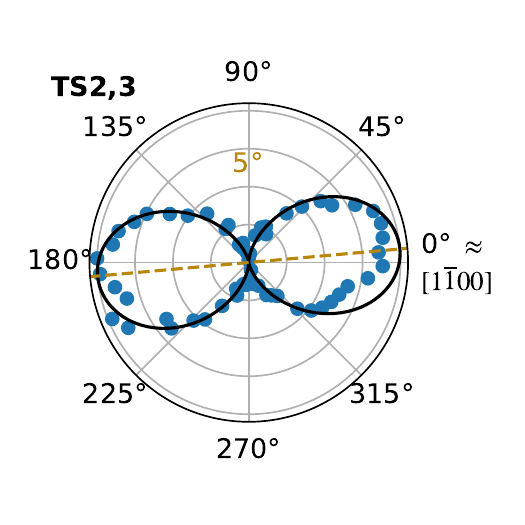}
    \end{subfigure}
\caption{(a) Confocal photoluminescence map integrated over the spectral range 767 to 774\,nm, suggesting two individual TS color centers (yellow) on an otherwise featureless background. (b) Spectra of position "d" with a sharp TS1 line and the TS2,3 double feature. For comparison, a background spectrum at position "b" and an ensemble spectrum recorded on a sample with higher TS concentration. (c)-(e). Analysis of the emission polarization of the TS1 line (upper row) and TS2,3 double feature (lower row) at three individual spots similar to "d". Distinct orientations can be assigned, rotated by $\approx$ 120°, in agreement with the crystal's symmetry. Dipole vectors can be assigned to next-nearest neighbor directions of the crystal structure with good accuracy.
}
\label{fig:spot_spectra_polarisation}
\end{figure*}

The photoluminescence experiment was performed at the C-face, because there a sparse distribution of TS centers is found. Figure \ref{fig:spot_spectra_polarisation}a displays a confocal PL map  
where for each pixel a full spectrum has been recorded; the color indicates the integrated intensity from 767\,nm to 774\,nm. Two pixels stick out with a clear signature of the TS1 line; all other pixels are considered as background luminescence. Figure \ref{fig:spot_spectra_polarisation}b shows the spectrum of pixel "d", in comparison with a background pixel "b". The spectrum displays the complete fingerprint of the ensemble spectrum (c.f. \cite{ruehl:18} and below) with a sharp ZPL and the weaker TS2,3 lines.

Looking at a larger number of bright spots with TS signature in an extended PL map, one finds that only three distinct polarizations occur. Examples of those are displayed in Figure \ref{fig:spot_spectra_polarisation}c-e (first row TS1, second row TS2,3).
Clearly an intensity dumbbell can be identified in the polar plot. Here, 0° corresponds to the crystal's $[1\overline{1}00]$-direction, however an inaccuracy due to experimental misalignment by a few degrees cannot be excluded. The three polarization signatures occur with a relative shift of 120°, as expected for the hexagonal crystal structure of 4H-SiC. Within the projection on the basal plane, the polarization vectors point towards nearest neighbors (Si-C).

Each of these defects displays both, the dominant TS1 and the weak TS2,3 double signature being strictly proportional in PL intensity within the given uncertainties. Hence, the single defect studies allow for the conclusion that the TS1, TS2 and TS3 originate from one and the same object \footnote{Respective $g^{(2)}$ measurements were performed but inconclusive due to the low count rate}. Moreover, their polarization dependence appears parallel to that of TS1, which indicates similarities in the dipole vector (despite spectrally different transitions). The overall findings support the notion that indeed individual color centers are observed.
When superimposing all three directions we find the very same isotropic polarization behavior observed in previous unperturbed ensemble studies \cite{ruehl:21}.
The unperturbed single defect studies come thus to the same result as ensemble splitting by strong perturbations.

\subsection{TS splitting under strain}

\begin{figure*}
    \centering
    \begin{subfigure}[b]{0.329\textwidth}
        \centering
        \includegraphics[width=\textwidth]{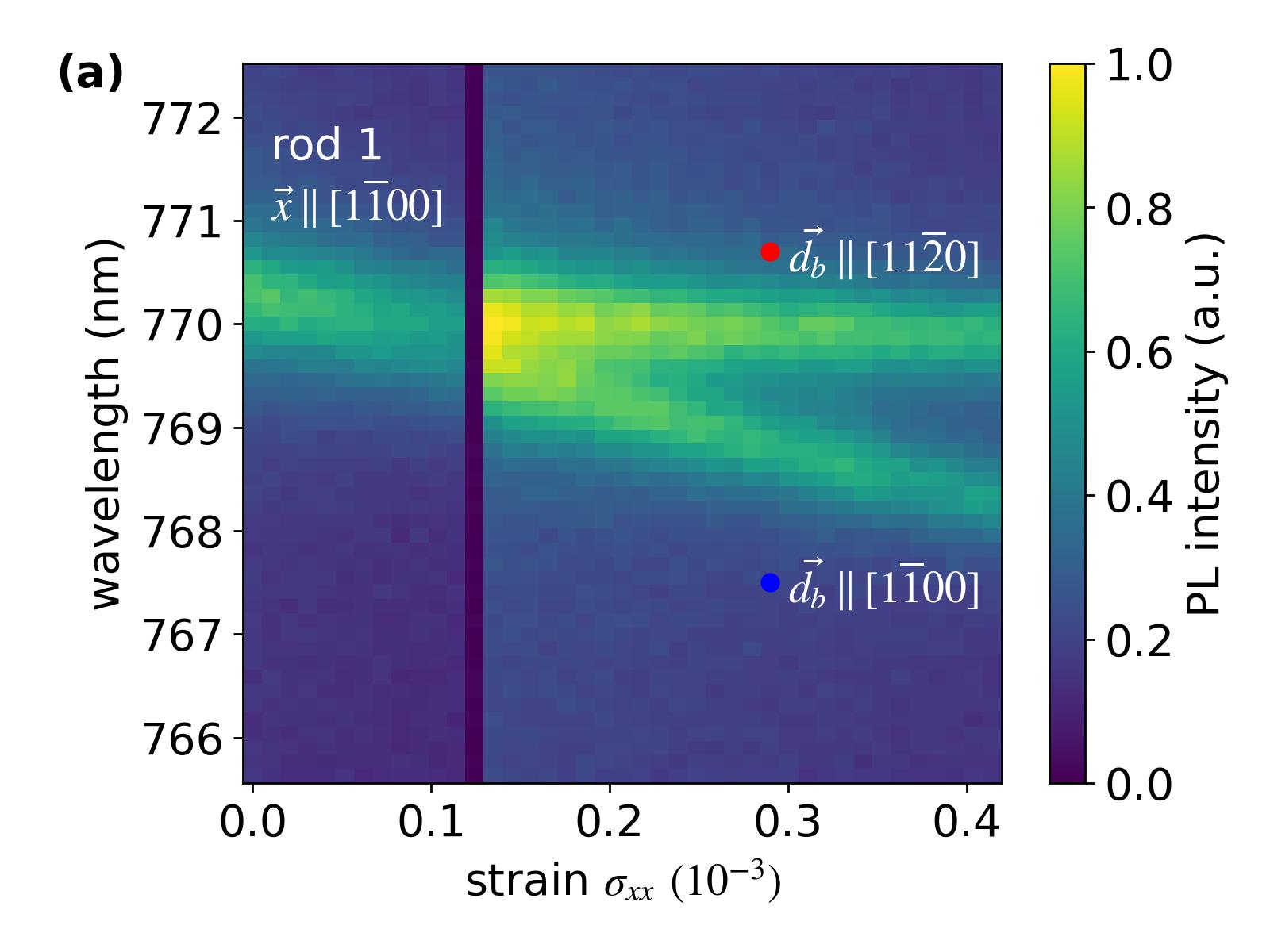}
    \end{subfigure}
    \hfill
    \begin{subfigure}[b]{0.329\textwidth}
        \centering
        \includegraphics[width=\textwidth]{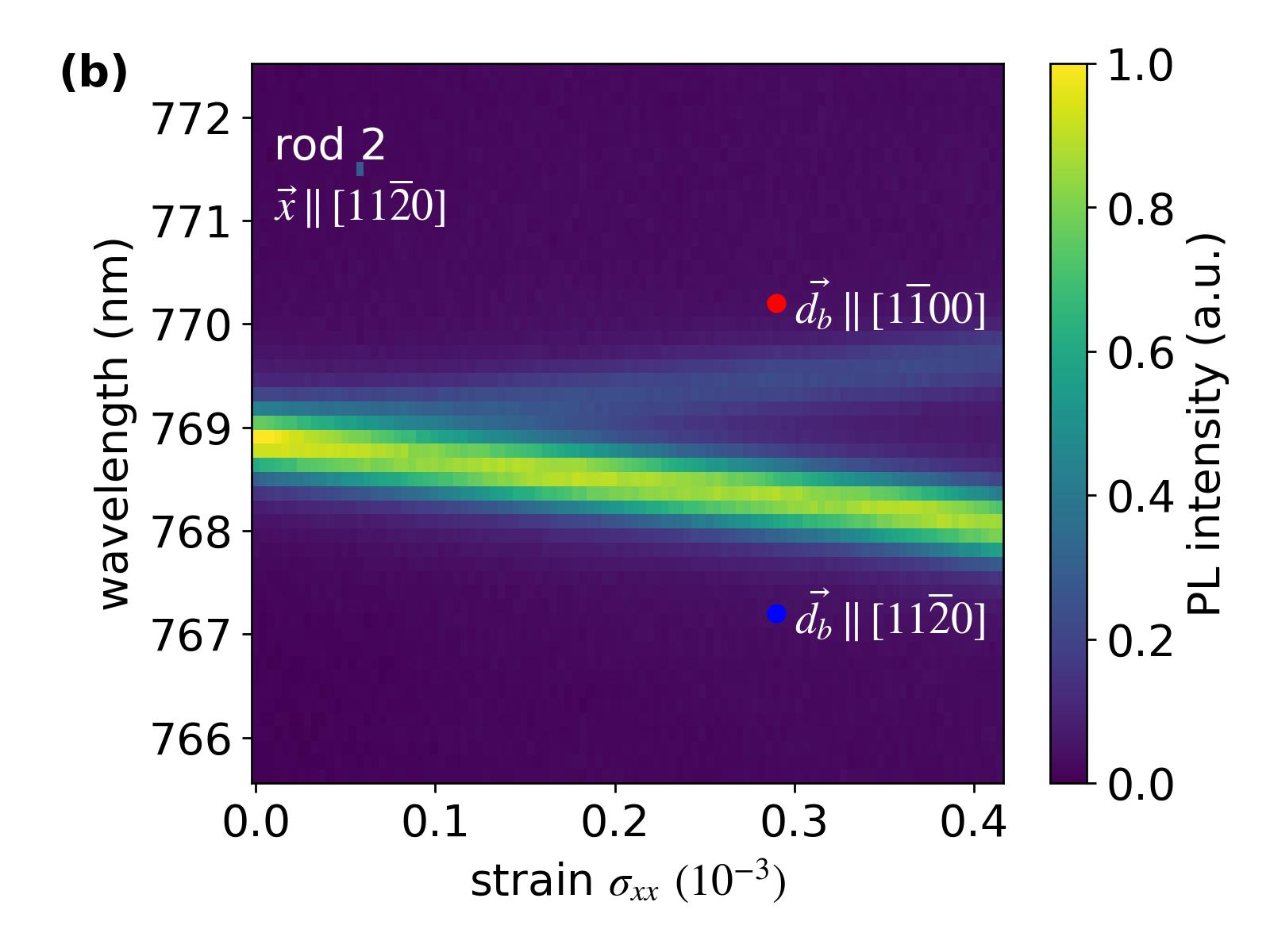}
    \end{subfigure}
    \hfill
    \begin{subfigure}[b]{0.329\textwidth}
        \centering
        \includegraphics[width=\textwidth]{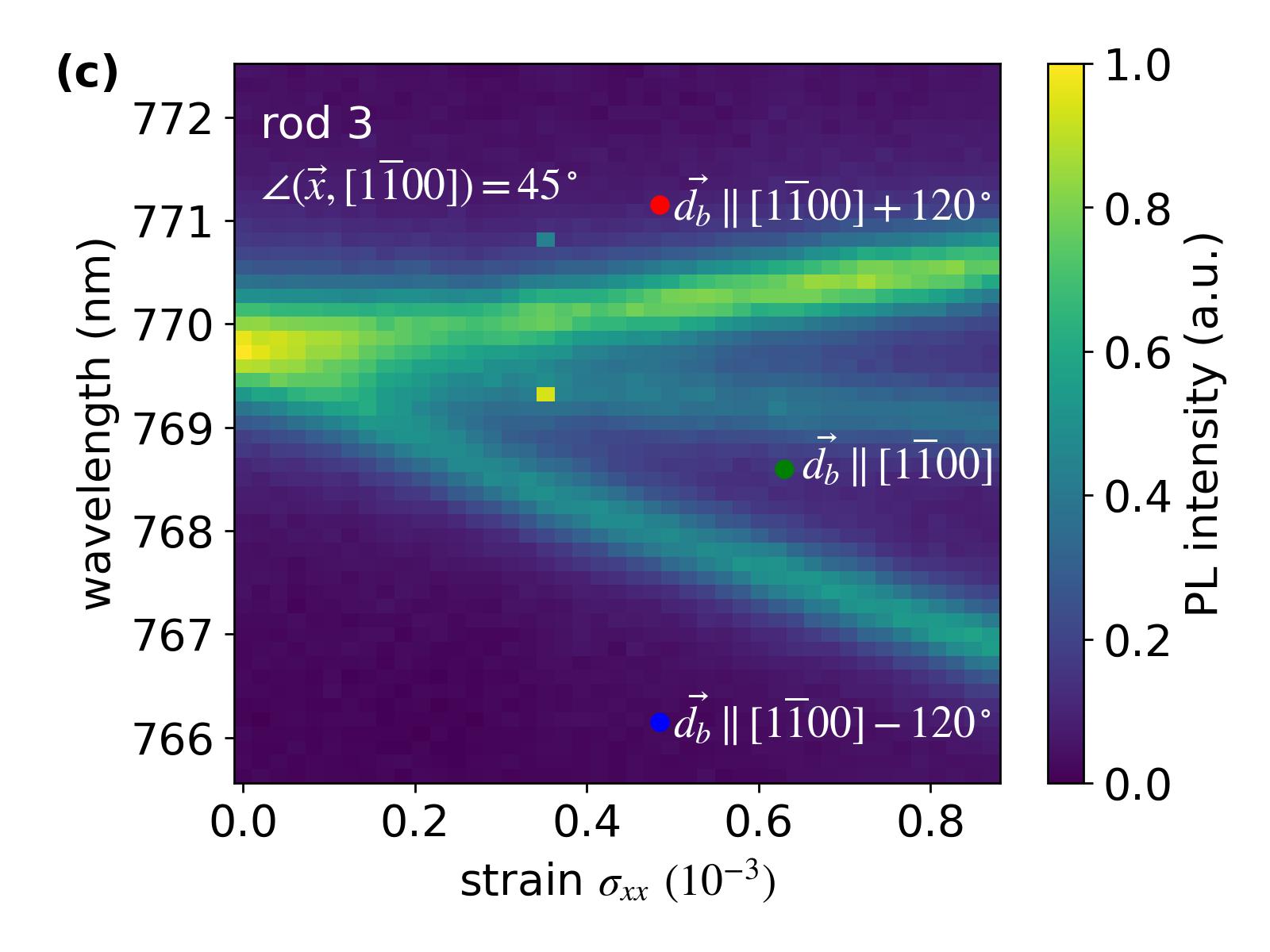}
    \end{subfigure}
    \\
    \begin{subfigure}[b]{0.32\textwidth}
        \centering
        \includegraphics[width=\textwidth]{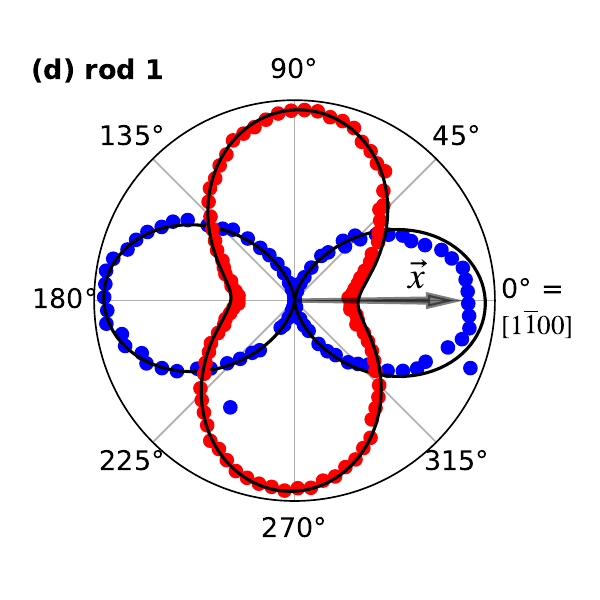}
    \end{subfigure}
    \hfill
    \begin{subfigure}[b]{0.32\textwidth}
        \centering
        \includegraphics[width=\textwidth]{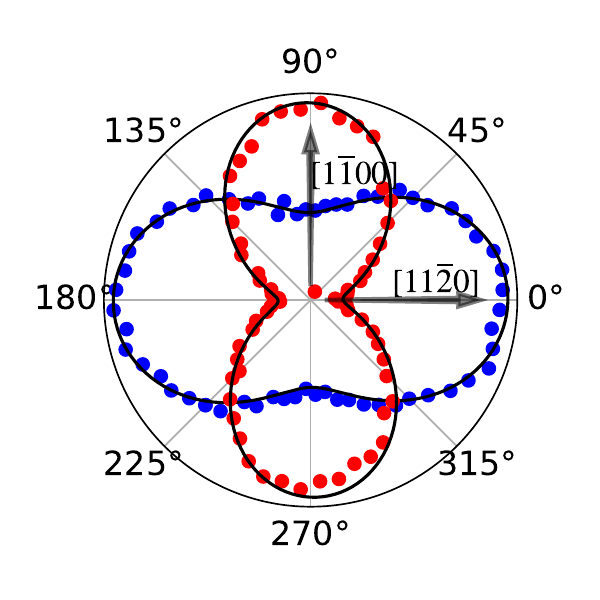}
    \end{subfigure}
    \hfill
    \begin{subfigure}[b]{0.32\textwidth}
        \centering
        \includegraphics[width=\textwidth]{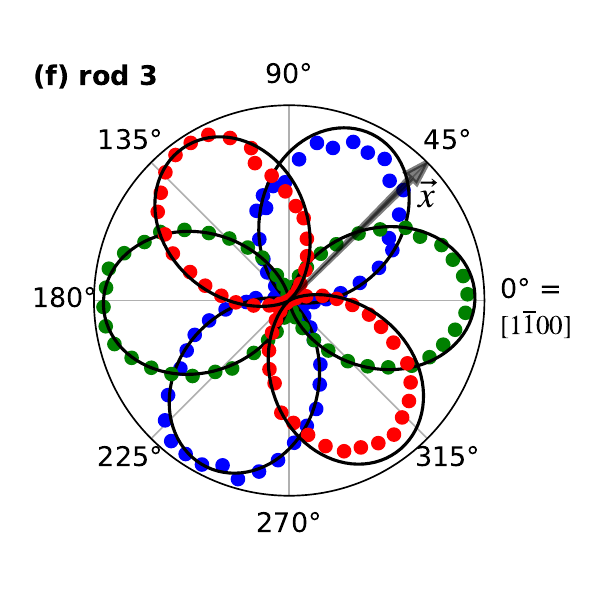}
    \end{subfigure}
    \caption{(a)-(c) Splitting of the TS1 line as a function of the applied strain along the rods' long side $x$. The predominant direction of strain with respect to the crystallographic axes differs for the different samples. The split lines are marked with the corresponding color and labeled with the direction of their dipole moment as derived from emission polarization of the single lines (d-f). Red dots represent the lines with longest wavelength, blue dots those with the shortest. The black lines are fits to the data, $\cos^2(\varphi + \varphi_0)$ for the lobes with pronounced waist and $\cos^2(\varphi + 120^\circ + \varphi_0) + \cos^2(\varphi - 120^\circ + \varphi_0)$ for the wider lobes in (d) and (e). In all cases $\varphi_0$ is interpreted as the basal orientation of the dipole moment $\vec{d_b}$.}
    \label{fig:strain_meas}
\end{figure*}

In order to investigate the color centers' response to strain, one has first to remind that strain is a tensorial quantity. When applying strain to a macroscopic sample, the color center may be affected by all tensor entries simultaneously. This is why we have opted for a rod along the $x$-direction with an aspect ratio of 6:1 between the clamping points, so the strain focuses on the $\sigma_{xx}$ component; all other tensor entries are at least one order of magnitude smaller (see SI). 
In order to vary $\sigma_{xx}$ with respect to the crystallographic axes, we cut three rods along $[1\overline{1}00]$ (basal projection of the Si-C bond), $[11\overline{2}0]$ (direction of C-C or equivalently Si-Si), and 45° in between. A mandrel exerted the force centrally, while the edges of the rods were fixed.
We performed PL spectroscopy on TS color center ensembles at the point of highest strain.
The results are shown in Figure \ref{fig:strain_meas}. One can clearly see that the TS1 line splits into two branches, with a linear dependence on the applied strain $\sigma_{xx}$, for rod 1 in $[1\overline{1}00]$ direction (Figure \ref{fig:strain_meas}a) and rod 2 in $[11\overline{2}0]$ direction (Figure \ref{fig:strain_meas}b). In the graph, we provide an assignment of the lines according to $\vec{d_b}$ (the projection of the transition dipole vector onto the basal plane, as derived from the emission polarization measurements displayed in Figure \ref{fig:strain_meas}d-f).
The relative intensity of the spectral lines has to be taken \textit{cum grano salis}, it depends on the excitation laser's polarization with respect to $\vec{d_b}$.

For rod 3, the result of which is displayed in Figure \ref{fig:strain_meas}c, we see a qualitatively different behavior: Here, we find a three-fold splitting.

For the three orientations, we can derive coefficients of the spectral response to strain which are displayed in Table \ref{tab:split_coeff}. From this, we can read off: The blue-shift is maximum when the dipole moment $\vec{d_b}$ is in line with the axis $\vec{x}$ of principal strain. In contrast, a maximum red-shift occurs with $\vec{d_b} \perp \vec{x}$. The detailed analysis is given in the SI.

\begin{table}
    \centering
    \caption{Coefficients of spectral response to strain, together with the angle between the respective lines' dipole moment $\vec{d_b}$ and the direction of the principal strain of the sample $\vec{x}$. The second values for $\vec{d_b} \parallel [11\overline{2}0]$ are the angles of the lines' two components.}
    \begin{tabular}{c|c|c|c}
         $\Delta \lambda / \sigma_{xx}$ (nm)& \multirow{2}{*}{$\vec{x} \parallel [1\overline{1}00]$} &  \multirow{2}{*}{$\vec{x} \parallel [11\overline{2}0]$} & \multirow{2}{*}{$\angle(\vec{x}, [1\overline{1}00]) = 45^\circ$}\\
 $\angle(\vec{d_b}, \vec{x})$& & &\\
         \hline
         \multirow{2}{*}{$\vec{d_b} \parallel [1\overline{1}00]$}&  -3.7&  1.6& -0.41\\
 & 0°& 90°&45°\\
 \hline
 \multirow{2}{*}{$\vec{d_b} \parallel [11\overline{2}0]$}& -0.63& -2.0&\multirow{2}{*}{$\diagup$}\\
 & 90° / $\pm$60°& 0° / $\pm$30°& \\
 \hline
         \multirow{2}{*}{$\vec{d_b} \parallel [1\overline{1}00] + 120^\circ$}& \multirow{2}{*}{$\diagup$} & \multirow{2}{*}{$\diagup$} & 0.96\\
 & & &75°\\
 \hline
         \multirow{2}{*}{$\vec{d_b} \parallel [1\overline{1}00] - 120^\circ$}& \multirow{2}{*}{$\diagup$} & \multirow{2}{*}{$\diagup$} & -2.8\\
 & & &15°\\
    \end{tabular}
    \label{tab:split_coeff}
\end{table}

The PL line with $\vec{d_b} \parallel [1\overline{1}00]$ displays a blue-shift for rod 1 (blue symbols) and a red-shift for rod 2 (red symbols). Both appear as a double lobe in the polar plot with a vanishing waist. Comparison with a sinusoidal behavior (black line, $\propto \cos^2 (\varphi)$) shows convincing agreement.
The other line ($\vec{d_b} \parallel [11\overline{2}0]$) observed in rods 1 and 2 appears a double lobe with a wider waist, which we interpret as a superposition of two sinusoidal contributions with $\cos^2 (\varphi + 120^\circ) + \cos^2 (\varphi - 120^\circ)$.

The three-fold splitting of rod 3 stems from the orientational mismatch between the three identified vectors $\vec{d_b}$ and the principal strain direction $\vec{x}$.
The consequence is a three-fold splitting.
The polarization dependence of all three lines has a vanishing waist.

The overall results are remarkably clear: Looking along the crystal's c-direction on the silicon-carbon tetrahedra, their three legs coincide with the observed polarization vectors, at least in the basal projection. This measurement, however, is insensitive to potential components in c-direction.


\subsection{Deep level transient spectroscopy}
In order to characterize a color center, its charge state is an important information, as it determines the interplay of the PL-active state and the electrochemical potential (Fermi level). For this purpose, we determine the charge transition levels.
The method of choice is deep level transient spectroscopy (DLTS) \cite{Lang1974} which is a time constant spectroscopy technique for the study of electron or hole emission from deep traps into the conduction or valence band respectively \cite{Kawahara2013, Bathen2019}.
By comparison with Shockley-Read-Hall statistics of defects, the energy position of charge transition levels and the electrical capture cross section can be determined.

\begin{figure*}
    \begin{minipage}{.49\textwidth}
        \centering
        \includegraphics[width=\linewidth]{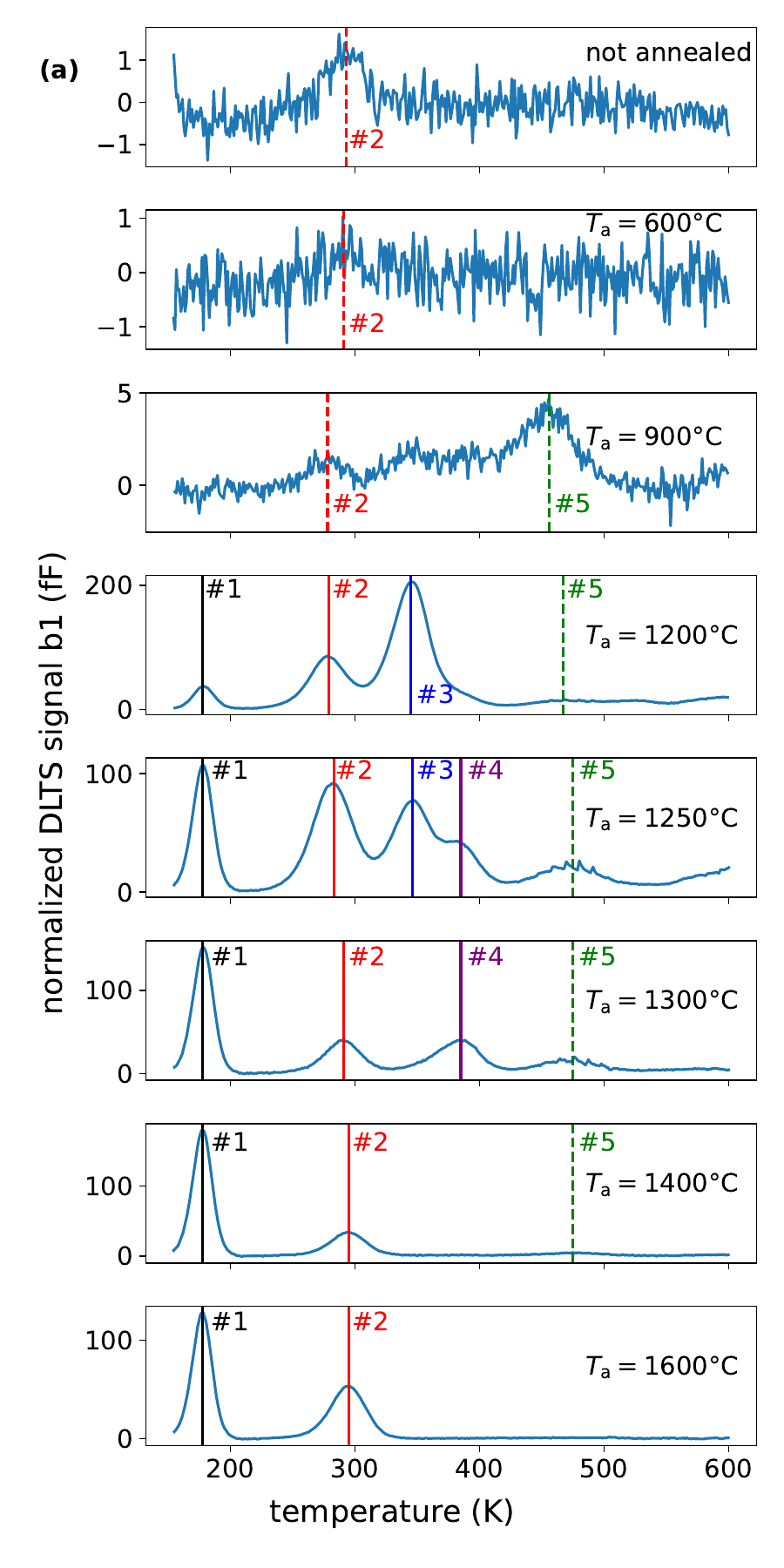}
    \end{minipage}
    \begin{minipage}{.49\textwidth}
        \centering
        \begin{subfigure}{\linewidth}
            \includegraphics[width=\linewidth]{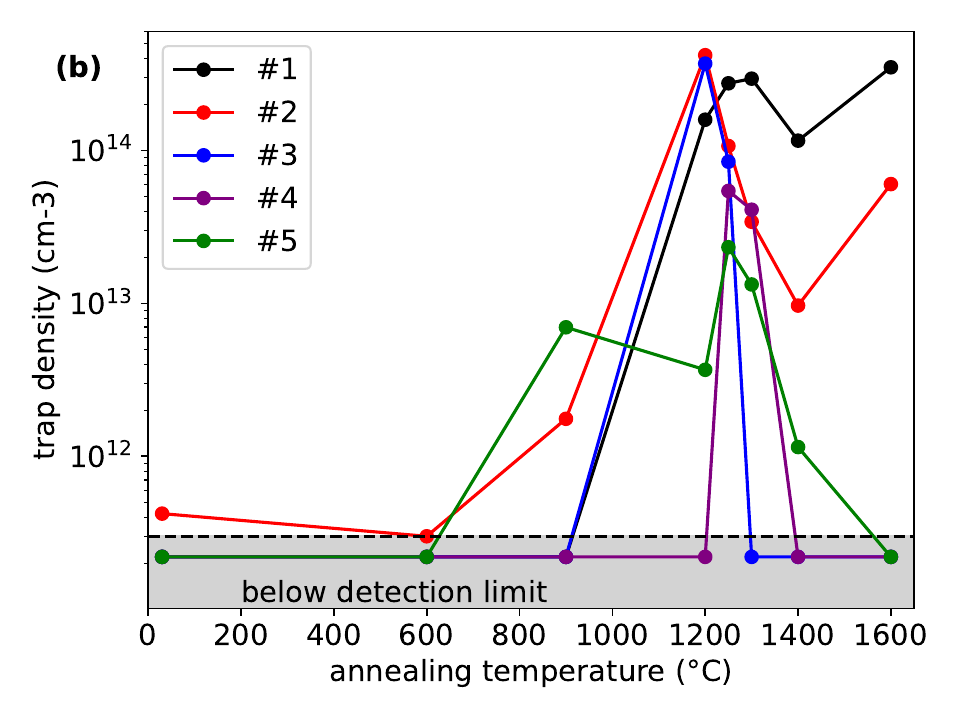}
        \end{subfigure}
        \begin{subfigure}{\linewidth}
            \includegraphics[width=\linewidth]{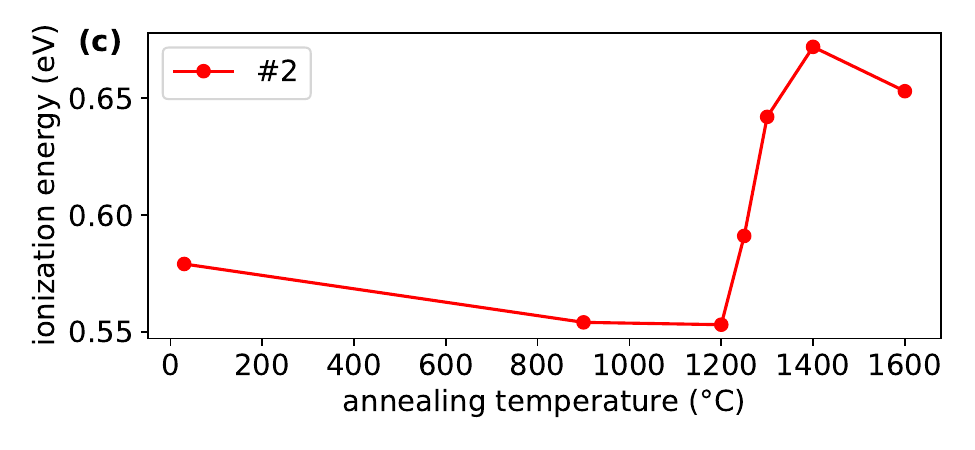}
        \end{subfigure}
        \begin{subfigure}{\linewidth}
            \includegraphics[width=\linewidth]{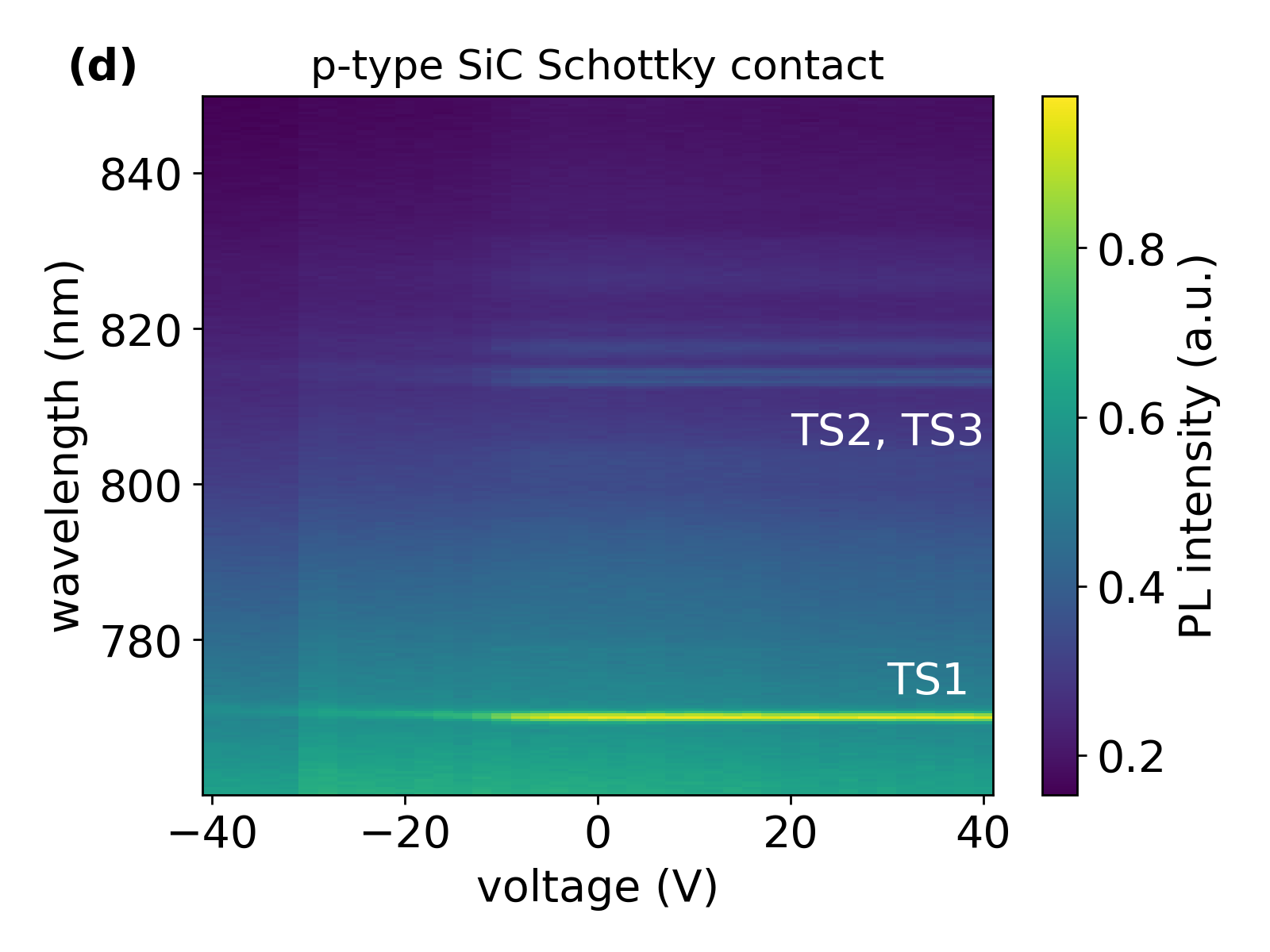}
        \end{subfigure}
    \end{minipage}
    \caption{\textbf{(a)} DLTS spectra of the investigated p-type samples at different annealing temperatures. For the spectra the b1 sine correlation function was used with a period width of 16\,ms and a pulse time of 10\,ms. The detected peaks are marked with a vertical line and numbered from \#1 to \#5. \textbf{(b)} The trap density of the different peaks as evaluated from DLTS. Most peaks are only detectable in a small range of annealing temperatures. The evolution the trap density belonging to peak \#2 is strikingly similar to the PL intensity reported by Rühl et al. for the TS defect \cite{ruehl:18}. \textbf{(c)} The activation energies measured for peak \#2. The energy shifts with increasing annealing temperature by more than 100\,meV. \textbf{(d)} PL measurement under a graphene-SiC p-type Schottky contact, adapted from \cite{schober:23}. The TS signature fades away towards negative voltages.}
    \label{fig:dlts}
\end{figure*}


We start our investigation in the lower part of the bandgap. For this purpose we move from semi-insulating to p-type material, in which the TS color center has previously been observed as well \cite{schober:23}. The DLTS spectra for different annealing temperatures are displayed in Figure \ref{fig:dlts}a.
The corresponding peaks in the various spectra are numbered from \#1 to \#5; their concentrations as a function of the annealing temperature is displayed in Figure \ref{fig:dlts}b. The occurrence of peaks \#1, \#3 and \#4 was reported in literature before \cite{Storasta2004, danno:07}. They and additionally the yet unreported defect \#5 reveal an annealing behavior unrelated to the one of the TS as observed in PL \cite{ruehl:18}.
The latter, however, strongly resembles the annealing behavior of peak \#2 (c.f. Figure \ref{fig:dlts}b).
Therefore, we tentatively assign this peak to the TS defect. From the Arrhenius evaluation of the DLTS spectra we obtain an ionization energy of approximately 0.6\,eV and a capture cross section in the range of $10^{-13}$ to $10^{-14}$\,cm$^2$.


A closer look to peak \#2 reveals a small shift of the peak position in the DLTS spectra with annealing temperature. This shift is also seen as a change of the ionization energy (c.f. Figure \ref{fig:dlts}c).
For a given defect, this is a rather unusual finding. For a tentative explanation of this effect, we direct the view on the statistical occurrence of the TS defect at various lattice sites (cubic, hexagonal) and various orientations, where rather subtle differences of the respective charge transition energies (even to the order of 100\,meV) should be expected \cite{Bockstedte2003}.
The DLTS method is not able to resolve their subtle differences in coexisting sub-ensembles. However, point defects at various lattice sites have also different formation energies. For this reason, different annealing conditions lead to different statistical weights in the sub-ensembles.
Hence we speculate that the shift of peak \#2 (c.f. Figure \ref{fig:dlts}a) and the associated ionization energy (c.f. Figure \ref{fig:dlts}c) result from a statistical redistribution during annealing.


A second series of DLTS experiments in the upper half of the bandgap did not reveal conclusive correlations with the appearance of the TS color center. The data are presented in the SI.

Altogether, correlated with the appearance of the TS defect we can identify a charge transition level 0.6\,eV above the valence band. When comparing this finding with previous studies at a p-type Schottky contact where the PL faded away under forward bias conditions (cf Figure \ref{fig:dlts}d \cite{schober:23}), we can assign the PL active state to the more negative charge state, i.e.\ the Fermi energy has to be above $E_\mathrm{V} + 0.6$\,eV.

\subsection{Absence of spin signals}
The spin of a color center is of interest, both for completing this thorough experimental characterization, and further it might be employed for quantum applications \cite{Atature2018}.

We investigated the TS color center with respect to its spin with electron spin resonance (ESR) and optically detected magnetic resonance (ODMR) \cite{Niethammer2016, Nagy2018, Wolfowicz2020, Castelletto2020}.
The applied proton irradiation results in a total number of $4 \cdot 10^{12}$ silicon vacancies according to SRIM calculations \cite{Ziegler2010}. It has been observed that the concentration of V$_\mathrm{Si}$ is proportional to the TS concentration evolving after annealing \cite{ruehl:18}, but the proportionality factor (yield) is unknown.
The sensitivity of the ESR measurement is $\approx 10^9$ spins. This means that we should be able to detect the TS defect's spin if the yield is as low as $10^{-3}$ per generated silicon vacancy.

The ESR measurements were performed with a Bruker EMX micro CW-ESR spectrometer at room temperature and at cryogenic temperature ($\approx 80$\,K). The applied microwave frequency was 9.85\,GHz with a Power of 1.8\,mW, while the magnetic field was scanned from 2500 to 4500\,G.
However, no ESR response was detected.
Note that all PL measurements were performed under laser illumination, but the ESR measurement was initially performed in the dark. This is why we repeated the ESR measurement under illumination with a 532\,nm laser at room temperature. Also this experiment gave no spin signal.

In order to cross check the ESR results, ODMR measurements were performed at 4\,K under illumination with a 635\,nm laser. The radio frequency was scanned from 40\,MHz to 4\,GHz. We could observe no response in the PL of the TS color center.
We conclude that we could not detect a spin signal. As it is generated out of silicon vacancies with an unknown yield we can conclude that either this yield is as low as $10^{-3}$ or the TS has no spin signature at all.

\section{\label{sec:sum}Conclusion}
We present extensive data of the physical properties of the TS defect in 4H-SiC, gained with single defect and ensemble studies.
The TS defect provides a stable and pronounced zero phonon line TS1 in low temperature photoluminescence experiments for Fermi energies above $E_\mathrm{V} + 0.6$\,eV. No spin signals could be observed.

The basal projection of the emission dipole vector is oriented along that of the Si-C bonds: $[1\overline{1}00]$, $[01\overline{1}0]$ and $[\overline{1}010]$.
From single defect investigations it can be concluded that each individual defect is associated with one out of these three orientations. Moreover the accompanying TS2 and TS3 lines stem from the very same object and share the same orientation.
The TS defect shows a strong response to mechanical strain (further to electric fields, see \cite{ruehl:21}). Our measurements calibrated the PL line shifts, which consequently result in a spectral splitting according to the three defect orientations.

Altogether, the TS defect provides a system where optical, mechanical and electrical degrees of freedom on the SiC platform can be coupled. The detailed parameterization also provides a benchmark for the still pending atomistic identification of the TS defect.

\begin{acknowledgments}
We acknowledge financial support by German Research Foundation (DFG, QuCoLiMa, SFB/TRR 306, Project No. 429529648). MS and MB received financial support from the Austrian Science Fund (FWF, grant I5195).
\end{acknowledgments}


\bibliography{references-TS}


\end{document}



\title{Supplementary Information: On the experimental properties of the TS defect in 4H-SiC}


\author{Johannes A.\ F.\ Lehmeyer\orcid{0000-0003-2041-9987}}%

 \affiliation{Lehrstuhl für Angewandte Physik, Friedrich-Alexander-Universität Erlangen-Nürnberg, Staudststr.\ 7, 91058 Erlangen, Germany.
}

\author{Alexander D.\ Fuchs\orcid{0000-0003-1896-9242}}%
 \affiliation{Lehrstuhl für Angewandte Physik, Friedrich-Alexander-Universität Erlangen-Nürnberg, Staudststr.\ 7, 91058 Erlangen, Germany.
}

\author{Zhengming Li}%
 \affiliation{Lehrstuhl für Angewandte Physik, Friedrich-Alexander-Universität Erlangen-Nürnberg, Staudststr.\ 7, 91058 Erlangen, Germany.
}

\author{Titus Bornträger}%
 \affiliation{Lehrstuhl für Angewandte Physik, Friedrich-Alexander-Universität Erlangen-Nürnberg, Staudststr.\ 7, 91058 Erlangen, Germany.
}

\author{Fabio Candolfi\orcid{0009-0004-1187-590X}}%
 \affiliation{Lehrstuhl für Angewandte Physik, Friedrich-Alexander-Universität Erlangen-Nürnberg, Staudststr.\ 7, 91058 Erlangen, Germany.
}

\author{Maximilian Schober\orcid{0000-0001-9788-3302}}
 \affiliation{Institute for Theoretical Physics, Johannes Kepler University Linz, Altenbergerstr. 69, A-4040 Linz,
Austria.}


\author{Marcus Fischer\orcid{0000-0001-6679-0509}}
 \affiliation{Erlangen Center for Interface Research and Catalysis (ECRC), Friedrich-Alexander-Universität Erlangen-Nürnberg, Egerlandtsr. 3, 91058 Erlangen, Germany.}

\author{Martin Hartmann\orcid{0000-0003-1156-6264}}
 \affiliation{Erlangen Center for Interface Research and Catalysis (ECRC), Friedrich-Alexander-Universität Erlangen-Nürnberg, Egerlandtsr. 3, 91058 Erlangen, Germany.}

\author{Elke Neu-Ruffing\orcid{0000-0003-1904-3206}}
 \affiliation{Rheinland-Pfälzische Technische Universität Kaiserslautern-Landau, Department of Physics, Erwin Schrödinger Strasse, D-67663 Kaiserslautern, Germany.}

\author{Michel Bockstedte\orcid{0000-0001-5720-4010}}
 \affiliation{Institute for Theoretical Physics, Johannes Kepler University Linz, Altenbergerstr. 69, A-4040 Linz,
Austria.}

\author{Michael Krieger\orcid{0000-0003-1480-9161}}%
 \affiliation{Lehrstuhl für Angewandte Physik, Friedrich-Alexander-Universität Erlangen-Nürnberg, Staudststr.\ 7, 91058 Erlangen, Germany.
}

\author{Heiko B.\ Weber\orcid{0000-0002-6403-9022}}%
\email{heiko.weber@fau.de}
 \affiliation{Lehrstuhl für Angewandte Physik, Friedrich-Alexander-Universität Erlangen-Nürnberg, Staudststr.\ 7, 91058 Erlangen, Germany.
}

\date{\today}

\maketitle

\section*{S1. Strain calibration}
The value of the main strain component $\sigma_{xx}$ is obtained in two steps. Using a force sensor, the force exerted by the mandrel was calibrated. The force is then converted to strain values using COMSOL Multiphysics simulations.

For a force of 1\,N, the strain tensor right atop the mandrel is calculated to be:
\begin{equation*}
\sigma = 
\begin{pmatrix}
6 & -0.015 & 0.23\\
-0.015 & -1.4 & -0.19\\
0.23 & -0.19 & -2.0
\end{pmatrix} \cdot 10^{-5}.
\end{equation*}
The off-diagonal entries of the strain tensor, the shear strain, are about one magnitude smaller than the diagonal, this is why we neglect them in further considerations.
Regarding the diagonal elements, the linear strain, $|\sigma_{xx}|$ is three to four times larger than $|\sigma_{yy}|$ and $|\sigma_{zz}|$. Thus we concentrate on $\sigma_{xx}$ in the main text as the predominant contribution to strain.

Note that all strain values are proportional to the applied force, which was varied from 0 up to 10\,N.

\section*{S2. Strain along the sample}

\begin{figure}[htbp!]
    \begin{subfigure}[b]{0.49\textwidth}
    \centering
    \includegraphics[width=\textwidth]{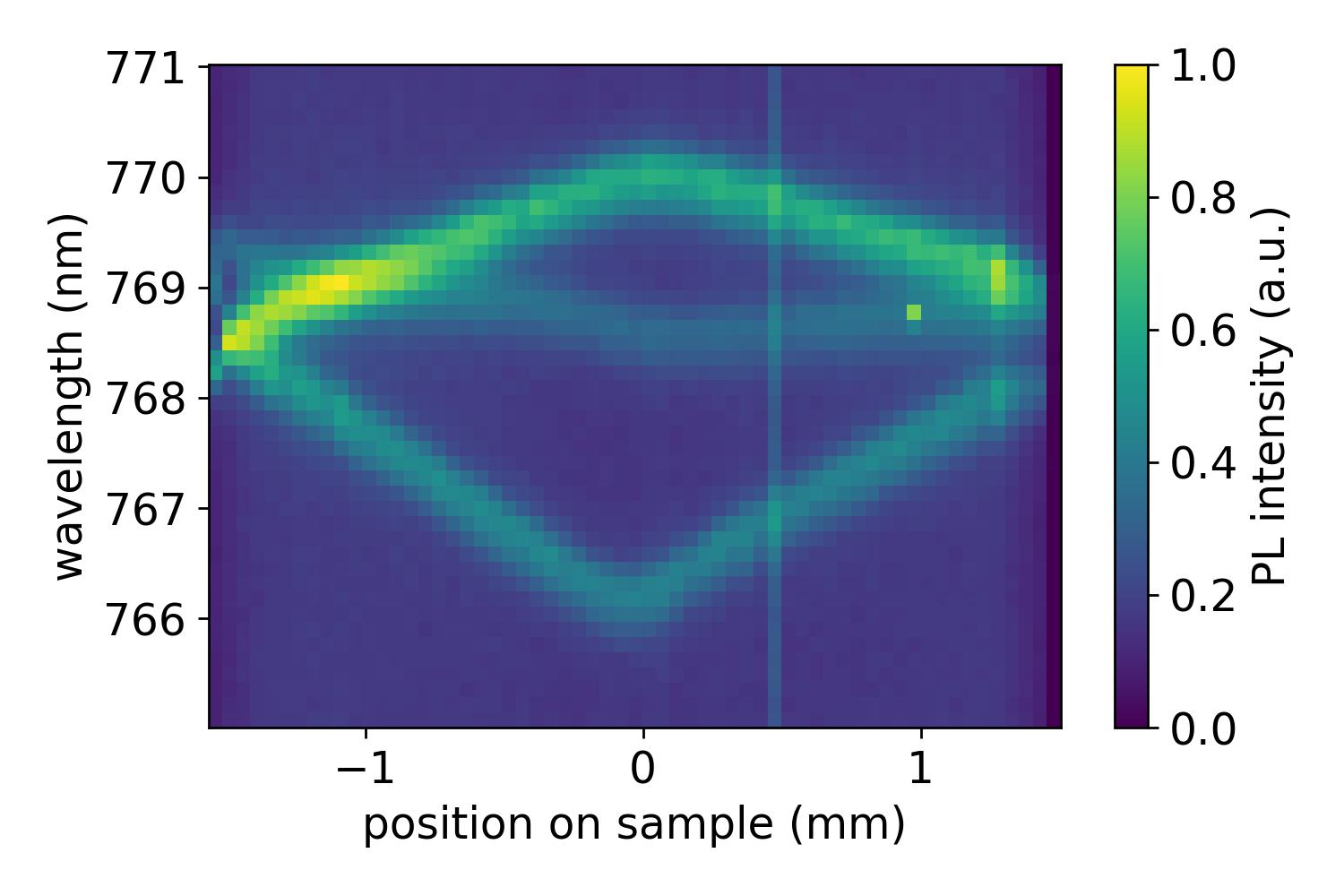}
    \caption{PL along sample}
    \end{subfigure}
    \\
    \begin{flushleft}
    \hspace{0.2cm}
    \begin{subfigure}[b]{0.4\textwidth}
    \includegraphics[width=\textwidth]{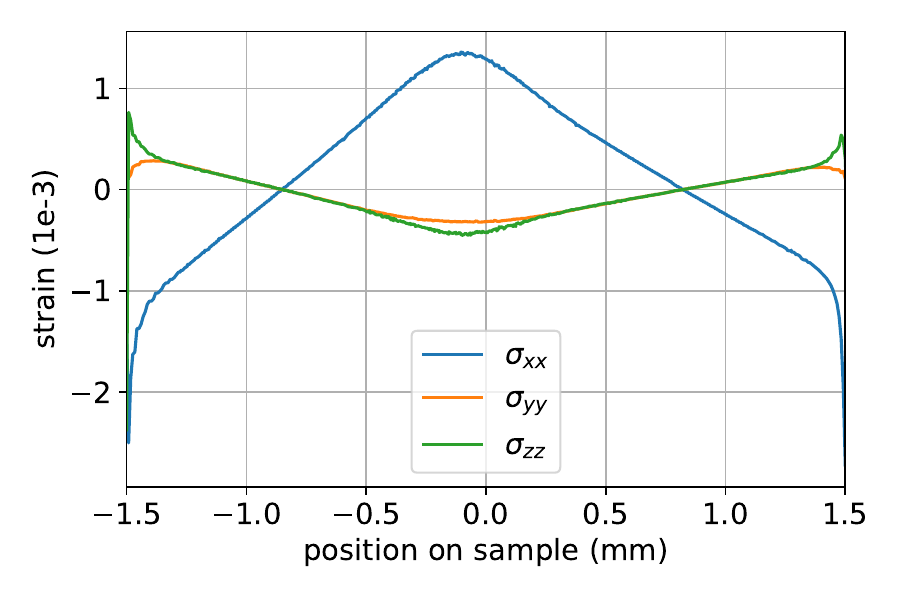}
    \caption{calculated strain along sample}    
    \end{subfigure}
    \end{flushleft}
    \caption{(a) Sweep along rod 3 where the long side of the sample $x$ forms a 45° angle with $[1\overline{1}00]$. The splitting is strongest around the middle of the sample, where the mandrel is pushing. (b) Strain as calculated with COMSOL. Because of small asymmetries of the setup, also an asymmetry in strain results.}
    \label{fig:strain_along_sample}
\end{figure}

In Figure \ref{fig:strain_along_sample}, we give a comparison of the photoluminescence of the TS1 line along rod 3 and the calculated strain along the sample. We selected rod 3 with $\angle (\vec{x},[1\overline{1}00]) = 45^\circ$ since there the three-fold splitting of the TS1 line becomes visible.
The splitting is strongest in the middle of the sample, where also the strain is highest. Towards the edges where the sample is clamped, the splitting decreases linearly, as does the strain. In Figure \ref{fig:strain_along_sample}a, the splitting reverses at the left edge of the sample. This is consistent with the strain $\sigma_{xx}$ becoming negative.

The asymmetry of the strain visible in Figure \ref{fig:strain_along_sample}b is due to the physical setup. The mandrel at the sample is operated by a lever, which gives a slight additional strain in one direction of the sample.
This asymmetry also becomes apparent in the splitting of the TS1 line (c.f. Figure \ref{fig:strain_along_sample}a). 

A qualitative mismatch occurs at the outer clamps: While we assumed perfect clamping in the simulation, the data rather show the behavior of simple support. This does only weakly affect the calibration.

\section*{S3. Influence of angle between strain and dipole moment}

\begin{figure}[htbp!]
    \centering
    \includegraphics[width=.49\textwidth]{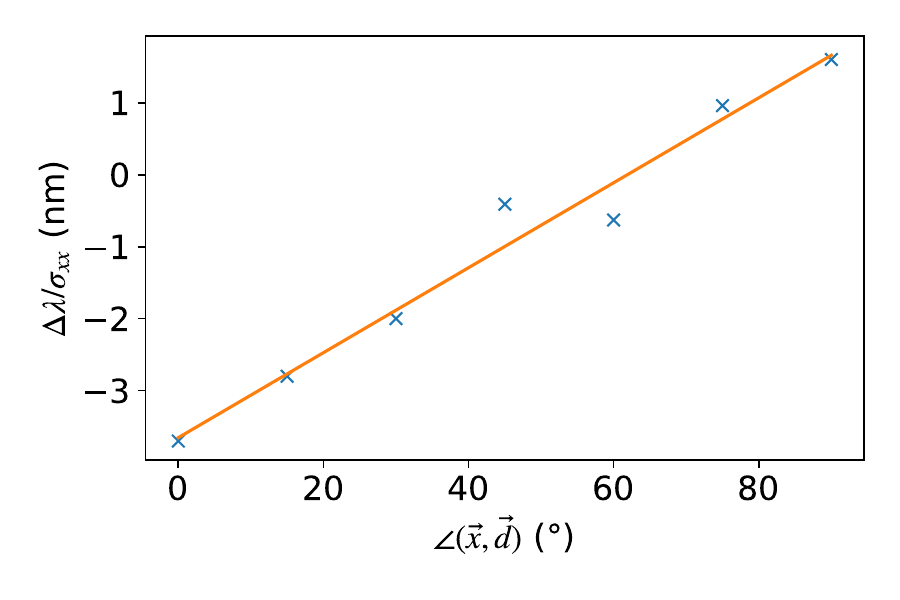}
    \caption{Spectral shift per strain versus the angle between the spectral line's dipole moment $\vec{d}$ and the direction of the principal strain $\vec{x}$.}
    \label{fig:slope_vs_angle}
\end{figure}

In Table I in the main text, we list the coefficients of spectral response to strain.
These coefficients were obtained by first fitting Gaussian peaks to the spectral lines in each spectrum and then performing linear fits to the peak positions as a function of strain.

In Figure \ref{fig:slope_vs_angle}, the coefficients are plotted against the angles between the dipole moment of the respective line and the direction of the principal strain. Except for the values at 45° and 60°, all values lie quite nicely on a linear relation.
Deviations may be because of the simplicity of the picture; we disregard the other strain components, since they are much smaller than $\sigma_{xx}$, but these components are still nonzero.

\section*{S4. DLTS measurements in n-type samples}

\begin{figure}[htbp!]
\includegraphics[width=.47\textwidth]{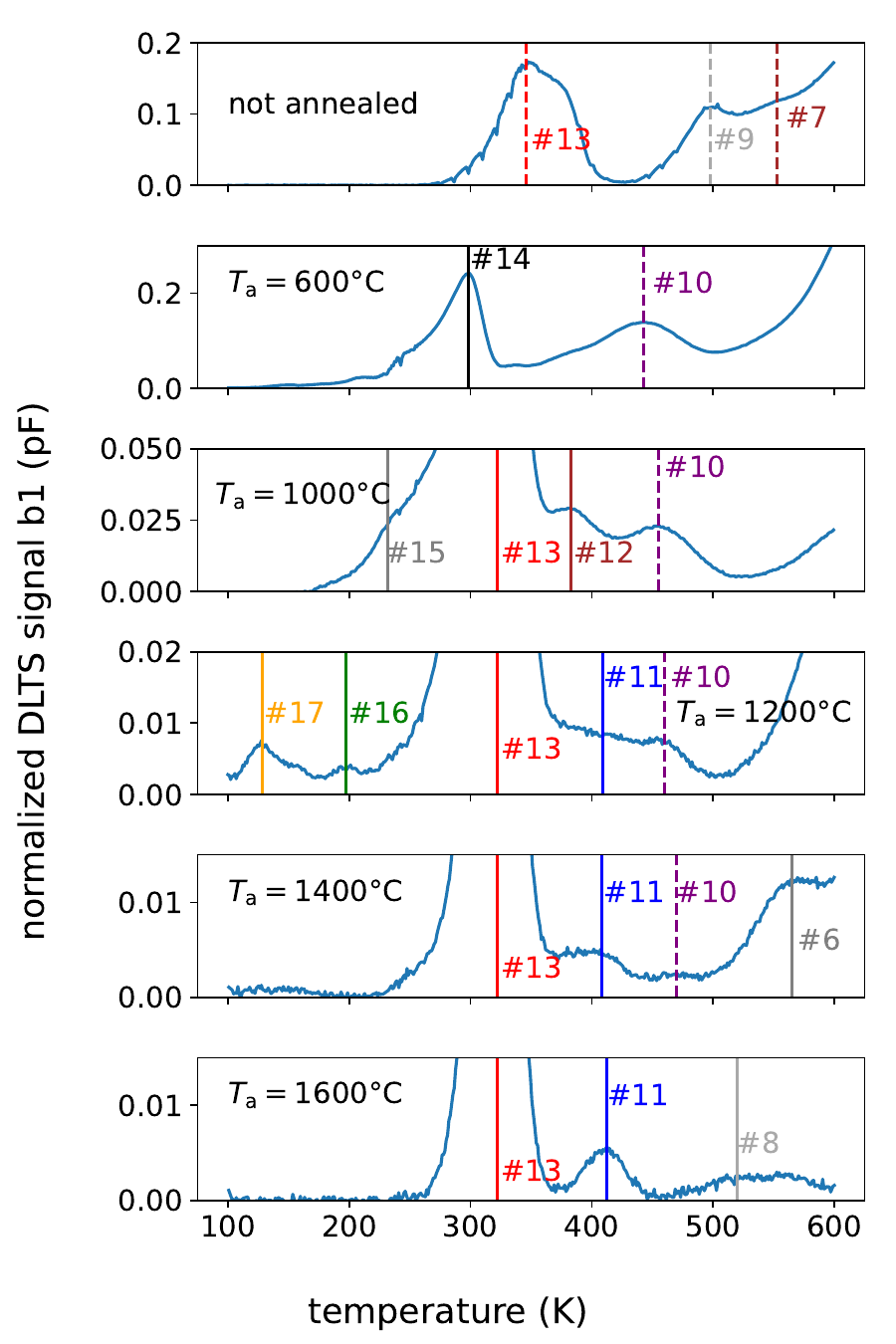}
\caption{\label{fig:dlts_n}DLTS spectra of the investigated n-type samples at different annealing temperatures. For the spectra the b1 sine correlation function was used with a period width of 16 ms and a pulse time of 10 ms. The detected peaks are marked with a vertical line and numbered from \#6 to \#17.}
\end{figure}

\begin{figure}[htbp!]
\includegraphics[width=.47\textwidth]{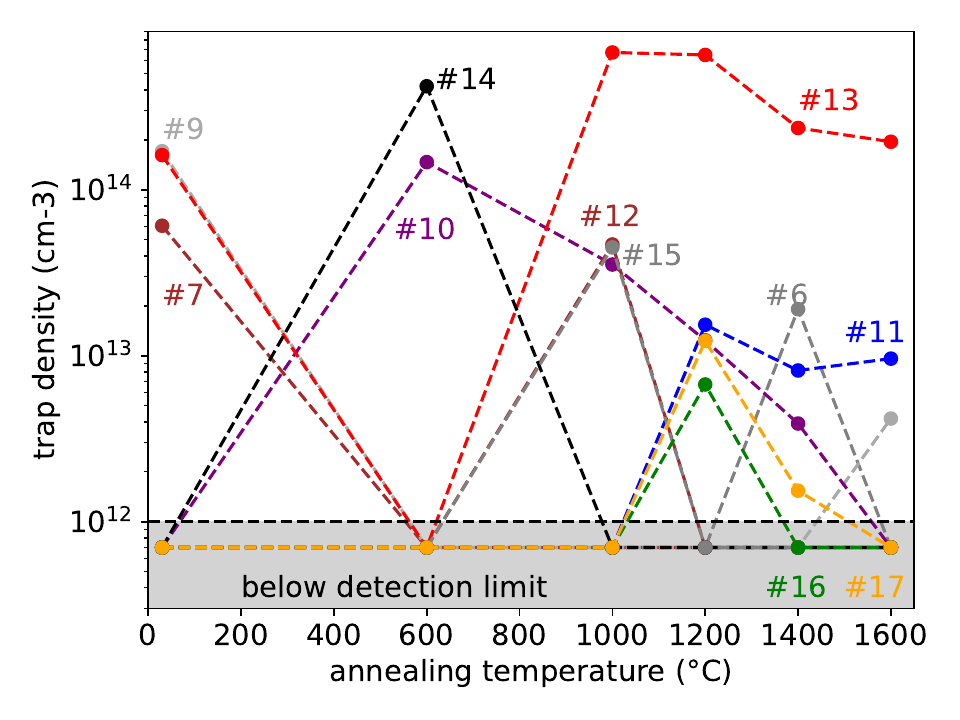}
\caption{\label{fig:dlts_n_density}The trap density of the different peaks in n-type samples as evaluated from DLTS. As expected, the Z$_{1/2}$ peak (\#13) shows the highest trap density and is mostly visible throughout all annealing temperatures.}
\end{figure}

\begin{figure}[htbp!]
\includegraphics[width=.47\textwidth]{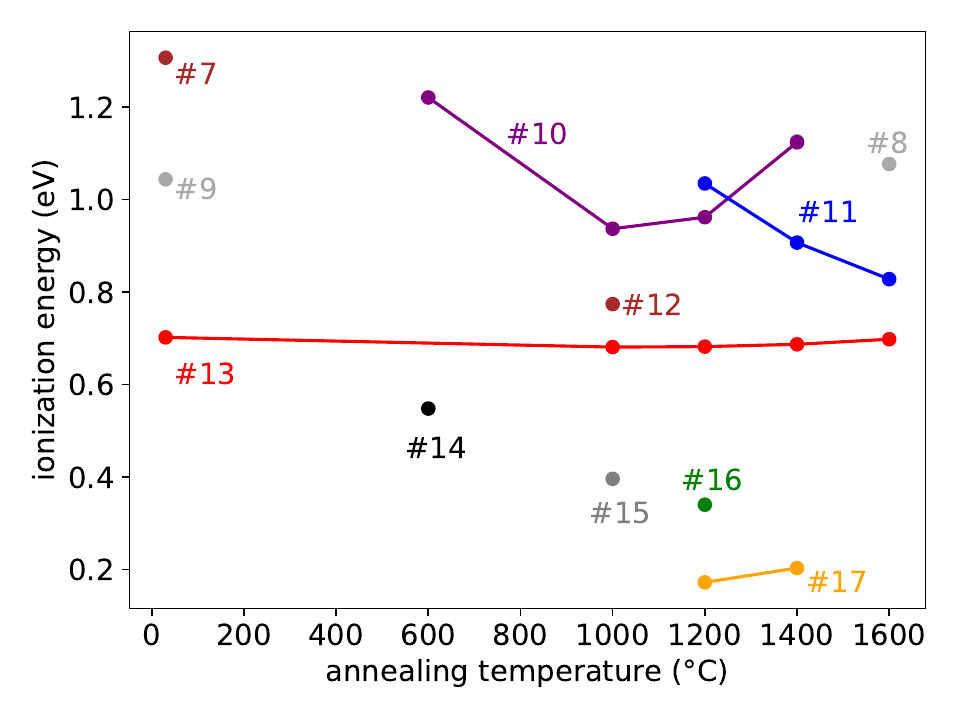}
\caption{\label{fig:dlts_n_activation}Ionization energies measured in n-type DLTS. Peak \#13 can be identified as the Z$_{1/2}$ peak.}
\end{figure}

\begin{figure}[htbp!]
\includegraphics[width=.47\textwidth]{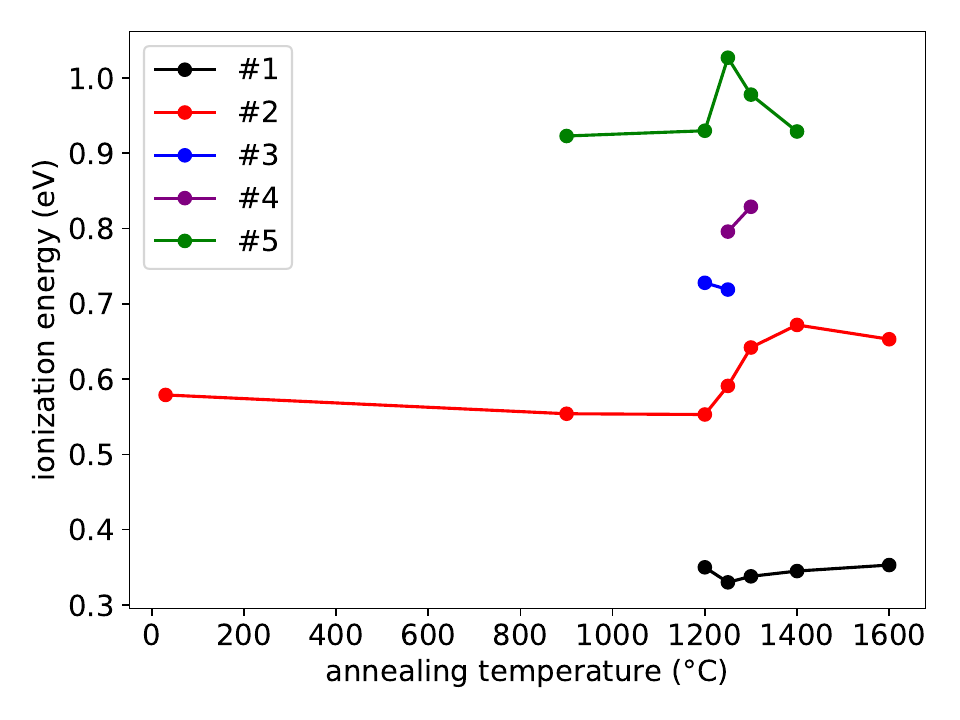}
\caption{\label{fig:dlts_p_activation}Ionization energies measured in p-type DLTS.
}
\end{figure}

For complementing the experimental picture, we take a look at the upper half of the bandgap by DLTS measurements on n-type Schottky contacts. Figure \ref{fig:dlts_n} displays the DLTS spectra obtained on proton irradiated samples after various annealing steps with temperatures up to 1600°C.
In contrast to p-type, the spectra reveal a multitude of overlapping peaks. The most dominant one corresponds to the well-known Z$_{1/2}$ defect which was previously identified as the carbon vacancy \cite{Kawahara2013}.

Figure \ref{fig:dlts_n_density} shows the trap densities of the observed defects. Other than for the p-type samples, we cannot identify a strong correlation with the PL signal of the TS defect.
Most defects appear at only a single or few annealing temperatures. The corresponding ionization energies are shown in Figure \ref{fig:dlts_n_activation}.
We conclude that we could not identify DLTS signatures that correlate with the TS defect in the upper half of the bandgap.

For completeness, the ionization energies of the other defects observed in p-type material are given in Figure \ref{fig:dlts_p_activation}.

\newpage
\bibliography{references-TS}